\documentclass[10pt, doublecolumn]{IEEEtran}
\usepackage{epsfig,latexsym}
\usepackage{float}
\usepackage{indentfirst}
\usepackage{amsmath}
\usepackage{bm}
\usepackage{amssymb}
\usepackage{times}
\usepackage{subfigure}
\usepackage{multirow}
\usepackage{psfrag}
\usepackage{hyperref}

\usepackage[table,dvipsnames]{xcolor}
\usepackage{booktabs,tabularx,array,makecell}
\usepackage{pifont}
\usepackage{fontawesome5}
\usepackage[most]{tcolorbox}
\newcommand{\cmark}{\textcolor{green!60!black}{\ding{51}}}
\newcommand{\xmark}{\textcolor{red!70!black}{\ding{55}}}

\newcolumntype{Y}{>{\raggedright\arraybackslash}X}

\usepackage{cite}
\usepackage{lastpage}
\usepackage{fancyhdr}
\usepackage{color}

\usepackage{amsthm}
\usepackage{bigints}
\newcounter{problem}
\newcounter{save@equation}
\newcounter{save@problem}
\makeatletter

\begin{document}
\title{Pinching Antennas Meet AI in Next-Generation Wireless Networks}

\author{ Fang Fang,~\IEEEmembership{Senior Member,~IEEE}, Zhiguo Ding,~\IEEEmembership{Fellow,~IEEE}, \\Victor C. M. Leung,~\IEEEmembership{Life Fellow,~IEEE}, and Lajos Hanzo, ~\IEEEmembership{Life Fellow,~IEEE} \vspace{-2em}   
\thanks{Fang Fang is with the Department of Electrical and Computer Engineering, and Fang Fang is also with the Department of Computer Science, Western University, London, ON N6A 3K7, Canada (e-mail: fang.fang@uwo.ca).}
\thanks{Zhiguo Ding is with the University of Manchester, Manchester, M1 9BB, U.K. (e-mail: zhiguo.ding@ieee.org).}
\thanks{Victor C. M. Leung is with the Department of Electrical and Computer Engineering, The University of British Columbia, Vancouver,
BC V6T 1Z4, Canada (e-mail: vleung@ieee.org).}
\thanks{Lajos Hanzo is with School of Electronics and Computer Science, University of Southampton, SO17 1BJ Southampton, U.K. (e-mail: hanzo@soton.ac.uk).}}
 \maketitle

\begin{abstract}
Next-generation (NG) wireless networks must embrace innate intelligence in support of demanding emerging applications, such as extended reality and autonomous systems, under ultra-reliable and low-latency requirements. Pinching antennas (PAs), a new flexible low-cost technology, can create line-of-sight links by dynamically activating small dielectric pinches along a waveguide on demand. As a compelling complement, artificial intelligence (AI) offers the intelligence needed to manage the complex control of PA activation positions and resource allocation in these dynamic environments. This article explores the `win-win' cooperation between AI and PAs: AI facilitates the adaptive optimization of PA activation positions along the waveguide, while PAs support edge AI tasks such as federated learning and over-the-air aggregation. We also discuss promising research directions including large language model-driven PA control frameworks, and how PA–AI integration can advance semantic communications, and integrated sensing and communication. This synergy paves the way for adaptive, resilient, and self-optimizing NG networks.
\end{abstract} 


\section{Introduction}

Next-generation (NG) wireless systems are expected to provide ultra-high data rates, massive connectivity, and ubiquitous intelligence. However, meeting these radical demands requires overcoming severe propagation losses and blockage for creating near line-of-sight (LoS) links. Recently, pinching antennas (PAs) have emerged as a flexible antenna technology for creating LoS links on demand \cite{PinchingDing2025_Tcom}. In particular, a PA is a small plastic clothespin attached to a dielectric waveguide, which locally perturbs the guided electromagnetic wave and causes a portion of it to leak into free space. By adjusting the position of the PA activated along the waveguide, the system can flexibly generate LoS links toward desired users. However, realizing the full potential of PAs in practice requires intelligent control to dynamically manage the activation of PAs using resource allocation in complex deployment scenarios. At the same time, emerging artificial intelligence (AI) aided applications, such as interactive extended reality (XR) and edge AI-aided Internet of Things (IoT), require ultra-low latency and highly reliable connections to support real-time data exchange and model updates. PAs constitute a promising solution for establishing strong LoS links. Integrating PAs with AI can combine the physical flexibility of PAs with the autonomous decision-making capabilities of AI, thus facilitating self-optimization of NG networks that can adapt to complex dynamic environments.

\begin{figure}
    \centering
    \includegraphics[width=0.5\textwidth]{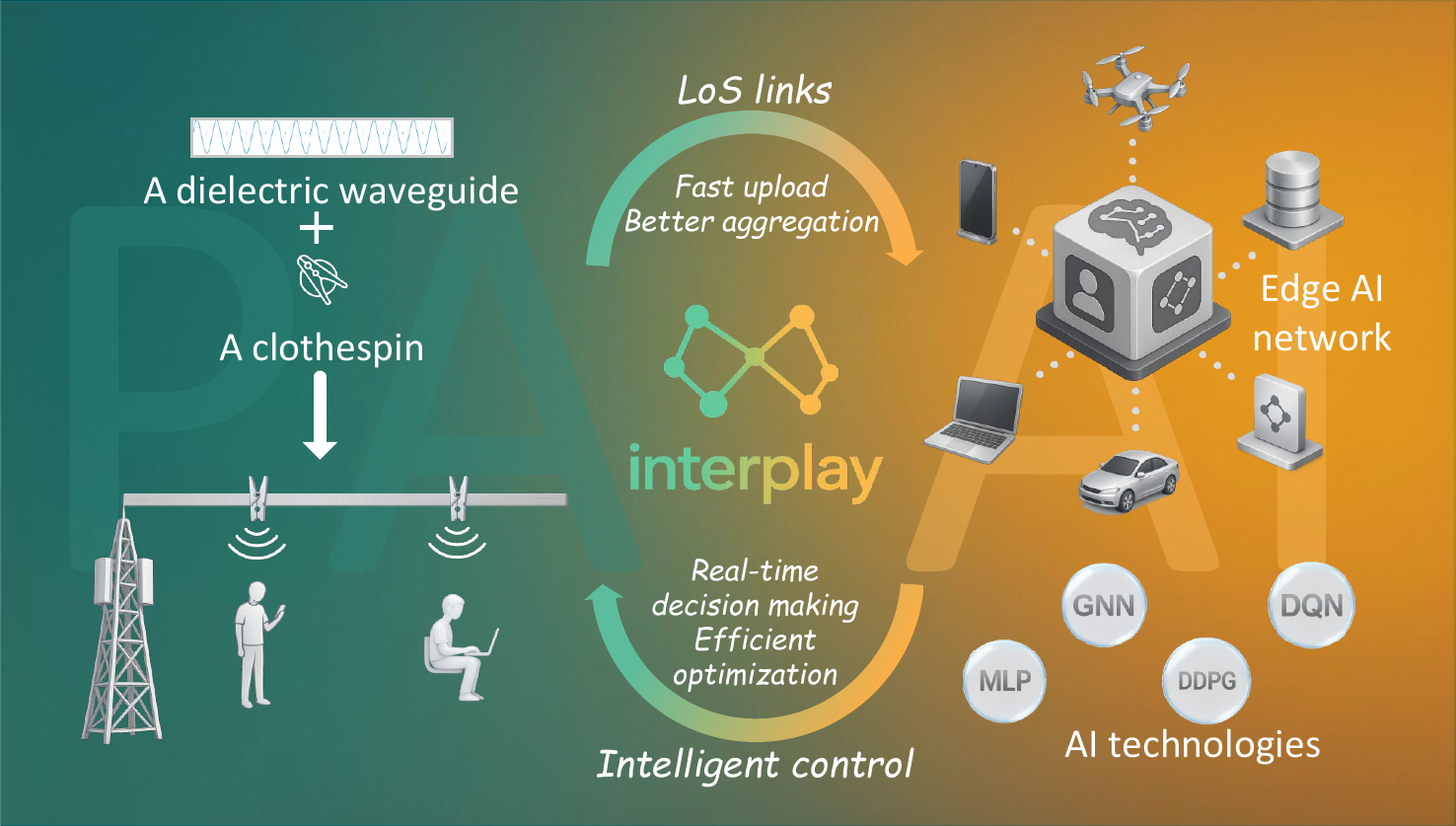}
    \caption{The interplay between PA and AI.}
    \vspace{-1.5em} 
    \label{PA AI interplay}
\end{figure}

A prototype of pinching-antenna systems was first demonstrated by NTT DoCoMo in 2022 \cite{NTTDocomo22}, where using small dielectric clips to ``pinch'' a low-cost waveguide and create radiation points. Hence PAs are capable of providing \textit{on-demand LoS connectivity} by flexibly activating the radiating point along the waveguide, under AI-control, as seen in Fig.~\ref{PA AI interplay}. Secondly, PAs offer \textit{attractive scalability}, since additional pinches can be deployed along the waveguide at negligible cost to create additional radiation points. Thirdly, PAs may mitigate the path loss and orchestrate strong LoS links, providing  `\textit{last-meter}' communication coverage \cite{PASSMagazine25}. 

While PAs offer remarkable physical flexibility, cost effectiveness and scalability, their practical activation relies on complex real-time decision-making problems that traditional optimization and rule-based techniques cannot efficiently solve. Tasks, such as determining \textit{where} and \textit{when} to activate pinching antennas, how to coordinate multiple PAs in tackling dynamic user demands and interference patterns, and how to promptly reconfigure the network in response to mobility or blockages, involve high-dimensional non-convex optimization in time-varying environments. AI provides powerful data-driven control mechanisms to efficiently tackle these challenges. By incorporating AI, networks can learn optimal PA configurations, predict environmental variations, and adapt to three-dimensional blockage conditions to maintain high-quality communication links. This synergy is particularly valuable in complex NG scenarios, where stable \textit{last-mile} connectivity and real-time decision-making are needed. With the power of AI, PAs can become intelligent components of a self-optimizing network infrastructure.

Similarly, PAs can effectively enhance the performance of AI-driven applications by establishing more reliable LoS links \cite{PassAirComp25} in data exchange. For example, in immersive XR environments, PAs can flexibly reconfigure indoor propagation to overcome LoS blockages, delivering the high data rates and stability needed for a seamless real-time user experience \cite{PassIndoor2025}. At the edge of the wireless network, PAs can improve distributed AI tasks; for instance, PAs improve wireless channels for over-the-air aggregation, reducing model misalignment in federated learning model updates \cite{BiboPAsFLTVT2025}. More broadly, by dynamically adjusting their positions, PAs are capable of directing capacity toward data-intensive nodes on demand, effectively supporting the low-latency, high-bandwidth data exchanges required by NG AI applications.

However, the `win-win' cooperation of PAs and AI hinges on challenges. The highly dynamic and unpredictable nature of AI-driven tele-traffic makes it difficult to coordinate PA reconfigurations in real time without incurring significant control overhead or additional latency. Multiple concurrent AI applications (e.g., several XR users or parallel learning tasks) may contend for limited PA resources, necessitating intelligent scheduling for avoiding interference and ensuring rate-fairness. Furthermore, the dynamic activation of PAs complicates traditional network operations. For example, channel estimation and beam training must account for on-demand antenna activation, which can significantly increase signaling overhead. These issues underscore the need for advanced algorithms and robust control frameworks to fully realize the potential of PA-AI cooperation in practice.

 Motivated by the above discussion, this article provides a concise overview of the emerging interplay between AI and PAs in NG wireless networks, as illustrated in Fig.~\ref{PA AI interplay}. Section II introduces the fundamentals and challenges of PAs and underscores the urgent need for PA-AI-cooperation. Section III explores how AI techniques can intelligently optimize PA activation in dynamic scenarios. Section IV discusses how PAs can in turn enhance AI performance. Section V outlines key challenges and promising future research directions. Finally, Section VI concludes the paper.

\section{AI for Pinching Antennas: Intelligent Deployment and Control}

\subsection{Fundamentals of PAs and Limitations}
As illustrated in Fig. \ref{PA AI interplay}, a pinching-antenna system (PAS) consists of a dielectric waveguide feeding a set of plastic clothespins, referred to as PAs. The PAs are placed at designated positions to form leaky waves from guided waves. By strategically activating PAs, the system can establish strong LoS links on demand. In particular, strategically activating a PA along the waveguide near a target user effectively transforms a weak or blocked link into a short, quasi-LoS hop. Naturally, physically relocating a PA along the waveguide is impractical. Hence a PA activation mechanism was proposed in \cite{Wangactive2025}, where multiple PAs are placed along the waveguide, each capable of being activated or deactivated. An active PA radiates electromagnetic waves into free space, while a deactivated PA remains electromagnetically transparent to the guided wave, facilitating near-instantaneous reconfiguration of PA activation positions.\par

Although pinching-antenna systems have attractive features, several challenges hinder the practical deployment of PAs. Firstly, the optimization for PA activation positions also known as PA activation design is inherently coupled with power allocation and beamforming, leading to a high-dimensional, non-convex joint optimization problem. Secondly, user mobility requires adaptive PA activation that follows user movement to sustain optimal link quality \footnote{User mobility is assumed to involve relatively slow movements (e.g., pedestrian speeds), and the distance between each user and his/her serving PA is moderate to ensure that PA reconfiguration can be accomplished within a feasible response time.}. Thirdly, indoor obstacles present a dual effect: while they may block LoS paths, they can also suppress co-channel interference \cite{BlockageDing25_CL}. Designing PA-activation strategies that exploit obstacles as constructive resources, rather than treating them solely as impairments, introduces additional complexity.

\subsection{AI enabled PAs}
\begin{table}[t]
\centering
\caption{Complexity comparison: search vs. learning approaches.}
\label{complexity}
\setlength{\tabcolsep}{3pt}
\renewcommand{\arraystretch}{1.05}
{\begin{tabular}{lcccc}
\toprule
Method & $N$ & $K$ & Evaluations & Time (1\,ms/eval) \\
\midrule
Brute-force & 20 & 6 & $20^6=6.4\!\times\!10^7$ & $\approx17.8$\,h \\
Brute-force & 30 & 4 & $30^4=8.1\!\times\!10^5$ & $\approx13.5$\,min \\
Brute-force & 30 & 8 & $30^8=6.6\!\times\!10^{11}$ & $\approx20.8$\,yr \\
Grid (3 passes) & 20 & 6 & $3\!\times\!6\!\times\!20=360$ & $0.36$\,s \\
Deep learning & -- & -- & $\approx1$ & $1$\,ms \\
\bottomrule
\end{tabular}}
 \vspace{-1.5em} 
\end{table}

\begin{table*}[!t]
\centering
\caption{Comparison of representative machine learning algorithms.}
\label{deep_networks}
\renewcommand{\arraystretch}{1.1}
\begin{tabular}{lccccc}
\toprule
\textbf{Algorithm} & 
\textbf{Time-varying} & 
\textbf{Control} & 
\textbf{Setting} & 
\textbf{Learning} & 
\textbf{Data Type} \\
\midrule
MLP & \xmark & Both & Centralized & Data-driven & Non-structural \\
DQN & \cmark & Discrete & Centralized & Data-driven & Non-structural \\
DDPG & \cmark & Continuous & Centralized & Data-driven & Non-structural \\
GNN & \cmark & Continuous & Centralized training, distributed execution & Data-driven & Structural \\
MADQN & \cmark & Discrete & Distributed training and execution & Data-driven & Non-structural \\
MADDPG & \cmark & Continuous & Centralized training, distributed execution & Data-driven & Non-structural \\
\bottomrule
\end{tabular}
\end{table*}

\begin{figure*}[!t]
    \centering
    \includegraphics[width=1\textwidth]{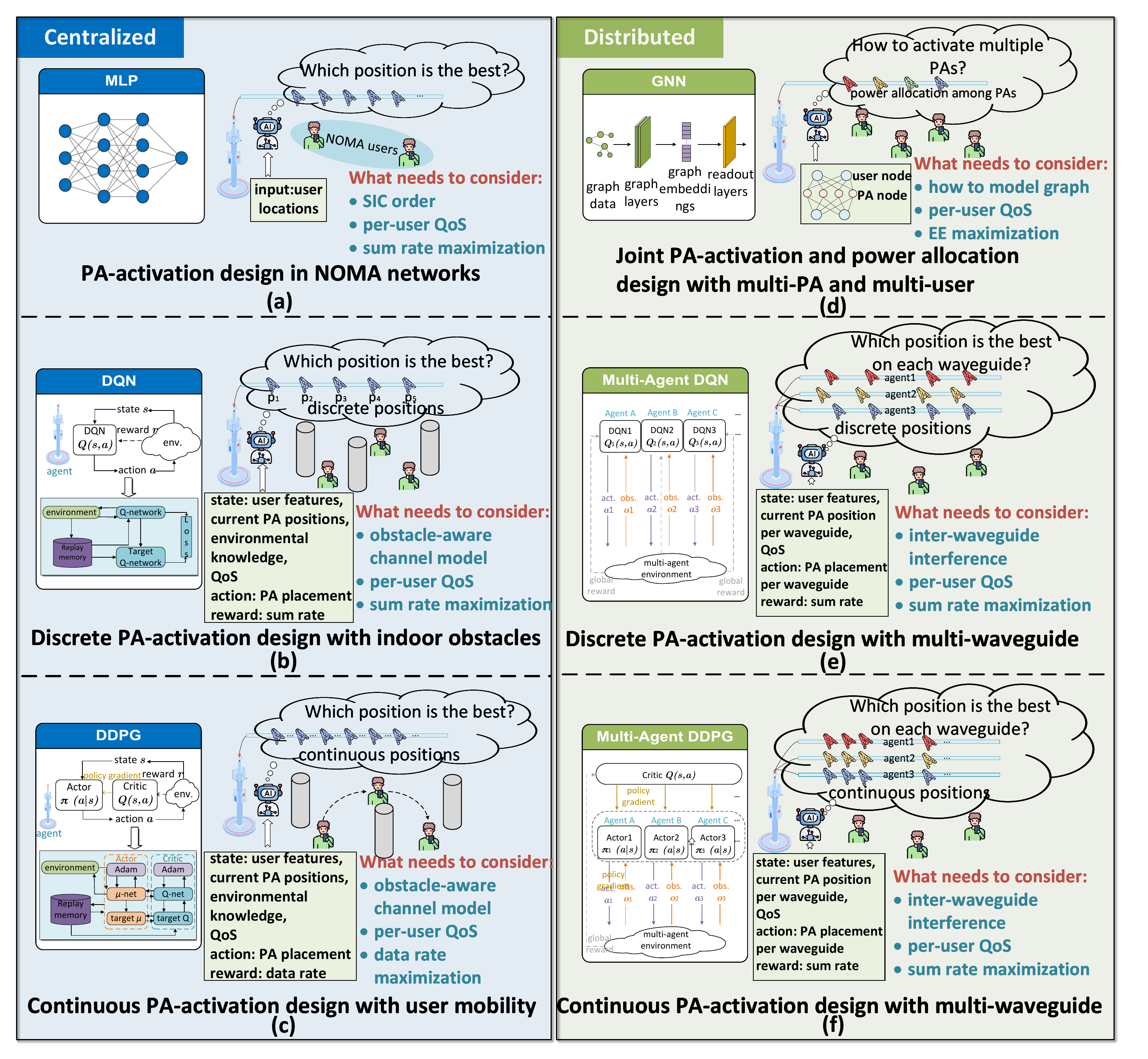}
    \caption{Overview of ML approaches for some PA-aided scenarios: centralized (MLP, DQN, DDPG), and distributed/multi-agent (GNN, MA-DQN, MA-DDPG), covering discrete/continuous activation positions, multi-waveguide coordination.}
    \label{overview of AI methods}
    \vspace{-0.5cm}
\end{figure*}
Conventional optimization approaches such as convex optimization, game-based, or matching-based methods are generally inefficient for PA control, as joint optimization often requires alternating updates over multiple variables, which causes latency and scalability issues. Moreover, obstacles lead to abrupt, non-smooth channel variations that further challenge these algorithms. Search-based methods, e.g., brute-force or grid search, can find near-optimal activation positions but have high computational costs. With $K$ waveguides and $N$ PA activation positions per waveguide, the exhaustive brute-force search has $\mathcal{O}(N^K)$ complexity while coordinate-based grid search has $\mathcal{O}(P K N)$ complexity, where $P$ denotes the number of iterative passes. To address these issues, AI-aided pinching-antenna systems learn the mapping from environmental states to optimal activation through offline training and perform fast online inference via a single forward pass, typically within milliseconds. A complexity comparison of search- and learning-based methods is summarized in Table \ref{complexity}.

\subsection{Promising Machine Learning Tools for PAs}
One of the most powerful tools of machine learning is deep neural networks (DNNs), which can efficiently infer a mapping from typical radio features, leading to beneficial actions. In the context of PA systems, a DNN processes inputs such as user locations, channel gains, 3D blockage information, and past PA activation positions. Its outputs are beamforming, power allocation, or PA-activation actions. Different types of DNNs are suitable for different optimization tasks, and these tools can be broadly categorized by their computational architectures. In centralized architectures, both training and inference are carried out by a controller relying on global network information. By contrast, distributed architectures support decentralized execution and/or collaborative training across PAs or user devices. The details are summarized along with representative tools in Table \ref{deep_networks}.\\
     \textit{Multi Layer Perceptron (MLP):} MLP is a multilayer feedforward model that learns complex nonlinear mappings from data via back propagation and gradient-based optimization. They are universal function approximators, widely used for regression and classification. Practical training relies on techniques such as normalization, dropout, weight decay, and early stopping for improving generalization.\\  
     \textit{Deep Q Networks (DQN):}  DQN is a value-based reinforcement learning (RL) method designed for discrete actions. It learns the action-value function using a deep net, stabilizes training by experience replay and a target network, while exploringly harnessing an \(\varepsilon\)-greedy policy.\\ 
     \textit{Multi-Agent DQN (MADQN):} MADQN extends value-based RL to multiple learners by involving discrete actions. A simple approach is independent DQN, where each agent learns its own Q-functions from local observations.\\
    \textit{Deep Deterministic Policy Gradient (DDPG):} DDPG is an actor–critic method conceived for continuous actions. An actor network outputs actions, a critic estimates the action-value function, and both use replay buffers and target networks; adding some noise to the actor’s output supports exploration diversity.\\
    \textit{Multi-Agent DDPG (MADDPG):} MADDPG is an actor–critic method configured for continuous actions. Each agent has a deterministic actor, while a centralized critic is used during training by exploiting the knowledge of conditions on the global state or joint observations and of all agents’ actions. At test time, each actor processes local observations only.\\   
    \textit{Graph Neural Networks (GNNs):} GNNs operate on graph-structured data through message passing: nodes aggregate information gleaned from neighbors and update their embeddings with the functions learned. This yields permutation-invariant models eminently suitable for node, edge, and graph level prediction.

\subsection{ML Approaches Across PA-aided Scenarios}
Fig. \ref{overview of AI methods} portrays ML techniques designed for different PA-aided scenarios. In each case, the inherent optimization challenge can be efficiently addressed by the corresponding ML approach as follows.\par

\textit{PA-activation design in NOMA networks}: As shown in Fig.~\ref{overview of AI methods}(a), optimizing PA activation in power-domain NOMA systems is challenging since the received power varies with PA activation position along the waveguide, which directly influences the optimal successive interference cancellation (SIC) order. This coupling renders conventional optimization computationally prohibitive. To overcome it, an MLP can learn the nonlinear mapping from user locations to near-optimal PA activation positions. Trained on simulated user distributions and channel realizations, the MLP rapidly predicts activation strategies that balance SIC order, user QoS, and overall sum-rate performance.

\textit{Discrete PA-activation design for indoor obstacles}: In indoor environments, PA activation must account for obstacles such as walls or pillars that cause abrupt channel variations. Even small user movements may block LoS links, which makes conventional gradient-based or convex optimization ineffective due to their smooth-function assumptions. Assuming PAs can only be activated at discrete candidate positions, the problem becomes a discrete decision-making task, where each activation is treated as an action. As a result, DQNs can learn obstacle-aware activation strategies through interactions with the environment. The state includes user features, current PA activation positions, environmental knowledge, and QoS requirements; the action selects a discrete activation position, and the reward is the achieved sum rate. Through training, the DQN develops policies that enhance throughput while maintaining reasonable fairness, even in complex indoor layouts\footnote{Sum-rate maximization may favor users with high-quality links over those with poor ones.}. An example is shown in Fig.~\ref{overview of AI methods}(b).

\textit{Continuous PA-activation design with user mobility}: Fig.\ref{overview of AI methods} (c) presents a case where PAs can be activated at any position along the waveguide. In contrast to discrete scenario, where candidate activation positions are fixed, the continuous case requires optimizing over an infinite set of possible antenna coordinates along the waveguide. Meanwhile, if users are mobile, PA-activation must adapt in real time to rapidly fluctuating channel conditions. In this dynamic setting, DDPG is well-suited because it directly handles continuous action spaces. The system state consists of user features, mobility patterns, current PA activation positions, environmental knowledge, and QoS requirements. The action is a continuous displacement of the PA along the waveguide, while the reward is defined by the data rate achieved. Through the actor–critic architecture, the agent learns policies that balance prompt adaptation with stable performance. \par

\textit{Joint PA-activation and power allocation design for multi-PA and multi-user scenarios}: \cite{xie2025graph} proposed maximizing energy efficiency in the multi-PA multi-user scenario of Fig.\ref{overview of AI methods} (d). The optimization challenge lies in jointly determining the antenna positions and power allocation, since activating a PA closer to a user can improve its link but simultaneously affect both interference and energy efficiency. This creates a tightly coupled and large-scale problem that is difficult to solve using conventional convex methods. To address this, the system can be modeled as a bipartite graph having user nodes and PA nodes, where channel relationships form the edges. Leveraging this structure, a GNN learns to output feasible PA coordinates and power levels that maximize energy efficiency under joint spacing and power constraints. \par

\textit{Discrete PA-activation design with multi-waveguide}: When multiple waveguides are harnessed as in Fig.\ref{overview of AI methods} (e), each equipped with a PA, the activation problem becomes more complex due to inter-waveguide interference. Selecting a position to activate a PA on one waveguide affects not only the users it serves but also the interference imposed on neighboring waveguides. This complexity creates a multi-agent decision-making environment, where independent optimization can lead to suboptimal or conflicting activations. To address this, the problem is modeled with each waveguide acting as an agent that selects from discrete candidate positions. A MADQN framework can be exploited, where agents interact with the shared environment and receive a global reward based on sum rate performance, while respecting per-user rate-constraints. By coordinating learning through centralized training but decentralized execution, the agents can infer activation strategies that effectively mitigate inter-waveguide interference and enhance the overall system throughput.

\textit{Continuous PA-activation design with multi-waveguide}: Fig.\ref{overview of AI methods} (f) illustrates a scenario, where multiple waveguides are deployed and PAs can be activated anywhere along their waveguides. A MADDPG framework is adopted, where each waveguide acts as an agent that outputs a continuous PA activation position.

\section{Pinching Antennas for Edge AI Networks}
\subsection{Fundamentals of Edge AI and Limitations}
In recent years, there has been a significant shift from traditional cloud-based AI toward edge AI, where edge networks process locally generated data and conduct AI-model training. The direct benefits of edge AI include alleviating the communication burden on centralized infrastructure, providing context-aware services and reducing response latency.

However, edge AI networks still face inherent challenges due to limited communication and computation resources at the edge, in the face of dynamic environments and mobility. To address these challenges, extensive research has focused on enhancing the efficiency of underlying edge networks, thereby enabling communication-efficient edge AI \cite{FL_Straggler}. Among the proposed solutions, improving the quality of wireless propagation has emerged as a potential approach. This includes harnessing advanced antenna technologies, such as fluid antennas and movable antennas, or `reconfiguring' wireless propagation environments using intelligent surfaces \cite{RIS_FL}. However, these techniques face inherent limitations: fluid and movable antennas are constrained by restricted mobility, while intelligent surfaces introduce additional control complexity and suffer from twin-hop double-pass loss. These drawbacks hinder their practical employment in edge AI systems.

\begin{figure}[!t]
    \centering
    \includegraphics[width=0.5\textwidth]{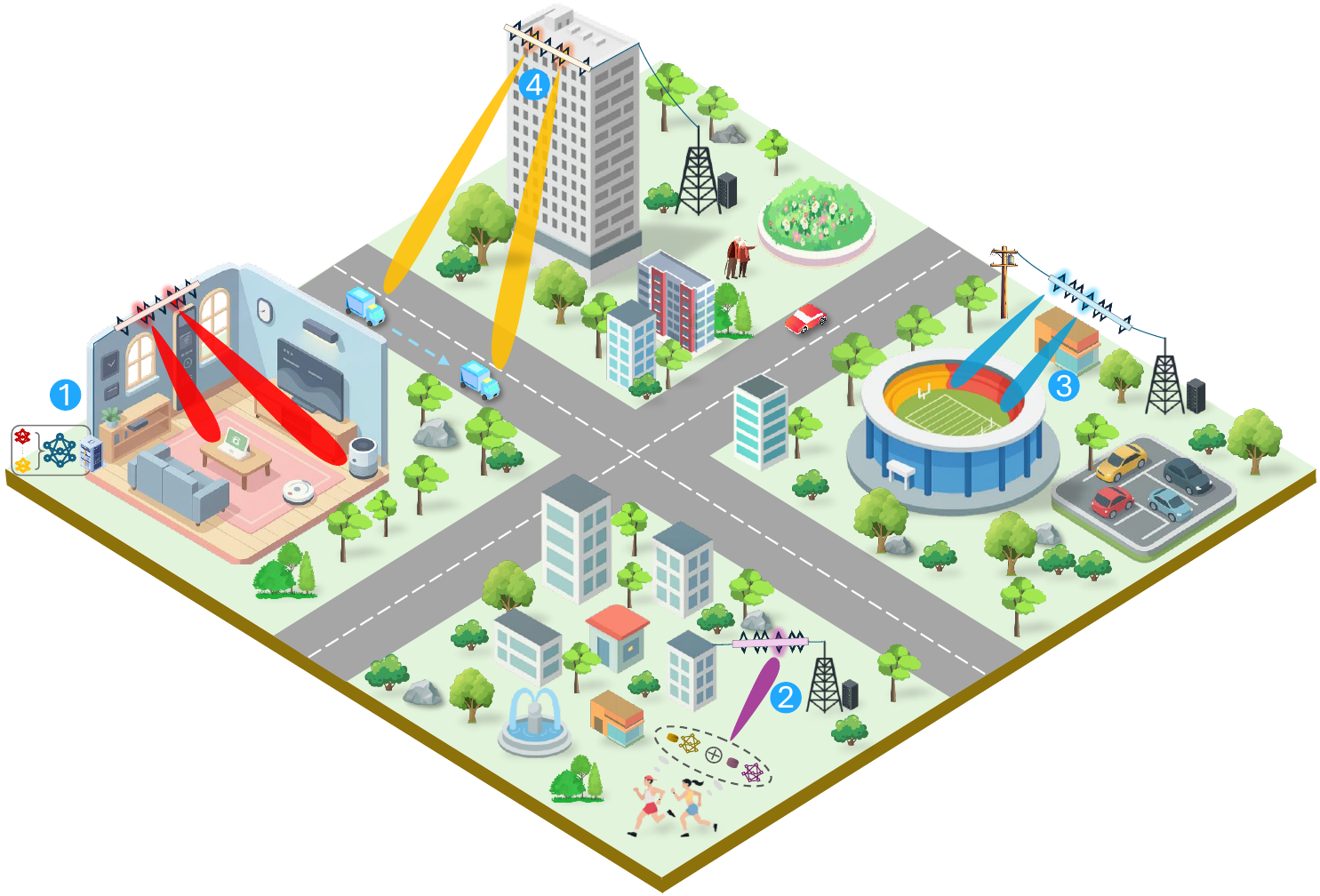}
    \caption{PAs for edge AI networks:
    \ding{172} PAs for indoor FL network;
    \ding{173} PAs for AirComp-enabled model update;
    \ding{174} PAs for real-time model training with data hotspot;
    \ding{175} PAs for real-time model training with mobile devices.} 
    \label{PA_for_AI}
    \vspace{-1em}
\end{figure}

\subsection{PA-enabled Edge AI Scenarios}
Motivated by the aforementioned challenges, in this section, we propose PA-assisted edge AI networks, where PAs are leveraged to enhance channel conditions and improve connectivity during edge AI training. By flexibly establishing LoS links at low cost, PAs offer an effective way of overcoming communication bottlenecks in dynamic and resource-constrained environments. We highlight four representative edge AI scenarios, as illustrated in Fig. \ref{PA_for_AI}: federated learning (FL) networks, over-the-air computation (AirComp)-enabled aggregation, dynamic-data edge AI and mobility-aware edge AI. 
These scenarios are not hierarchical but complementary: FL networks support the distributed training paradigm; AirComp emphasizes communication-efficient model aggregation; dynamic-data networks reflect challenges in handling evolving heterogeneous datasets; while mobility-aware networks focus on device dynamics and link stability. Together, they provide a holistic view of the diverse bottlenecks in edge intelligence. 
By integrating PAs into these representative frameworks, we demonstrate their potential as a key enabler for adaptive, resilient, and scalable edge AI systems.

\textit{Supporting reliable and balanced links for FL}:
As a promising distributed learning paradigm in edge AI, FL has attracted considerable attention from both academia and industry due to its advantages in preserving data privacy and reducing communication overhead. In FL, raw data remain on local devices, and only model parameters are exchanged between the edge server and devices to collaboratively update a global model. Consequently, wireless networks serve as the critical backbone that enables frequent and reliable server–device interactions. 
However, the dynamic nature of wireless channels and the scarcity of communication resources often create bottlenecks that lead to the well-known ``straggler issue''. In particular, devices with limited resources or poor communication conditions significantly delay the overall training process, leading to slow convergence and reduced training efficiency \cite{FL_Straggler}.

\begin{figure}
    \centering
    \includegraphics[width=3in]{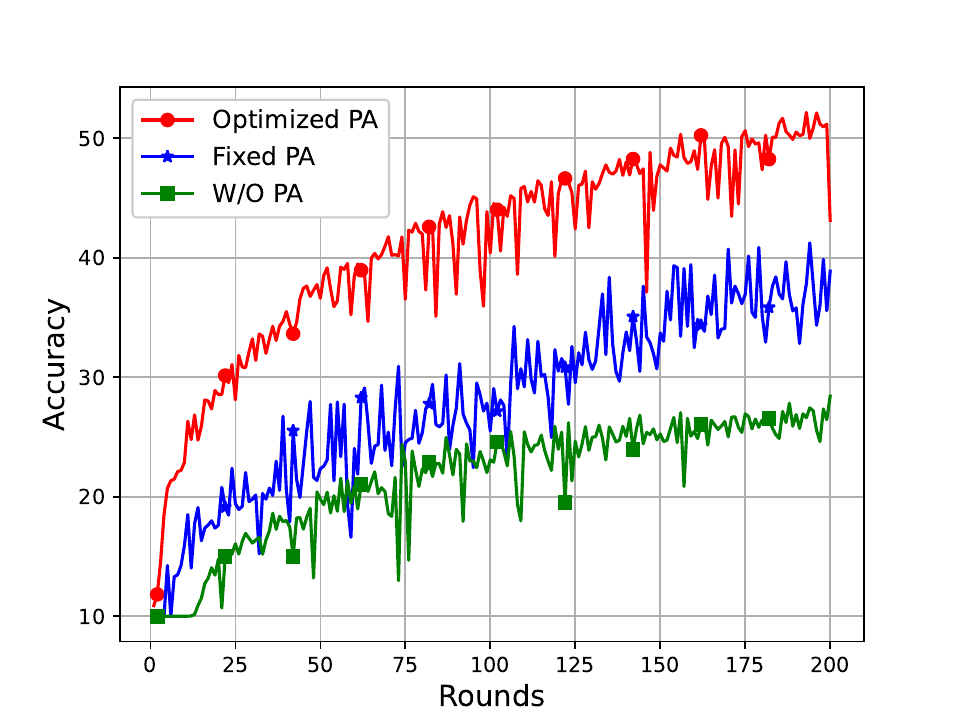}
    \caption{Performance of PA-assisted FL system.}
    \vspace{-2em}
    \label{Pin_FL}
\end{figure}

To overcome communication bottlenecks in FL, we propose to harness PAs to provide a scalable, cost-effective solution by adaptively establishing LoS links through PA-activation along dielectric waveguides. 
In conventional FL frameworks, devices with valuable local data may still become stragglers due to limited communication resources or unfavorable channel conditions, thereby prolonging the overall training time. Traditional schemes often discard these stragglers to accelerate convergence, but this inevitably sacrifices model accuracy, as their unique data remain unexploited—particularly under non-IID data distributions.
By incorporating PAs, these high-value devices can upload their local model parameters for global aggregation through optimized PA links. As a result, high-rate LoS transmission significantly enhances the overall performance of FL for such devices.
As shown in Fig. \ref{Pin_FL}, compared with the scheme operating without PA (W/O PA), a fixed-position PA activation already enhances FL performance by improving the communication conditions of specific stragglers. Further performance gains are achieved through optimized PA activation, where the activation position is adaptively determined to accelerate the FL training process. The superiority of this optimized scheme validates the effectiveness of PA-assisted edge intelligence.
Therefore, optimizing PA activation proves highly beneficial for improving FL performance, particularly in resource-limited and straggler-prone environments.

However, integrating PAs into FL systems also introduces new challenges. Firstly, the definition of stragglers in conventional FL has to be revisited. Instead of considering only wireless channel conditions, straggler identification should jointly account for both data value and communication quality to strike an improved balance. Although our previous work \cite{BiboPAsFLTVT2025} proposed a novel fuzzy logic-based device classification scheme, further research is needed to develop more efficient designs that incorporate additional device characteristics and system-level factors.
Moreover, the coexistence of pinching and conventional antennas within the underlying FL network requires careful design. In particular, advanced interference mitigation techniques and effective clustering strategies may be necessary to ensure high system performance.

\textit{Enhancing AirComp for communication-efficient model updates}:
Beyond the straggler issue of edge AI networks, model aggregation also represents a critical bottleneck. Due to the limited wireless communication bandwidth, the efficiency of local update aggregation is severely constrained, thereby limiting the overall system performance.
To address this challenge, AirComp has emerged as a powerful technique for wireless data aggregation in edge learning systems by exploiting the superposition property of multi-access wireless channels \cite{AirCom_FL}. In this way, multiple edge devices can simultaneously upload their local updates over shared wireless resources, thereby improving communication efficiency and reducing latency for global model updates. Unlike conventional edge learning systems that follow a compute-then-communicate paradigm, AirComp enables a compute-while-communicating paradigm, which significantly enhances the efficiency of model aggregation.

Despite its advantages, AirComp-aided edge learning suffers from aggregation errors caused by channel fading and signal misalignment. These errors accumulate over training rounds, slowing convergence and biasing the final global model. In particular, devices with poor communication conditions further exacerbate the error propagation, severely undermining learning performance. Therefore, enhancing the reliability of AirComp through improved wireless link quality is essential for unlocking its full potential in edge AI.

PAs offer a promising pathway toward this goal \cite{PassAirComp25}. By adaptively establishing LoS links through flexible activation along dielectric waveguides, PAs can substantially enhance channel quality and reduce signal misalignment in AirComp transmissions. This integration provides a scalable, low-cost approach to boosting aggregation reliability and accelerating convergence.
Within this integrated framework, a key challenge lies in minimizing aggregation error, while jointly optimizing PA activation and resource allocation across edge devices. The problem is inherently non-convex and tightly coupled, posing significant difficulties for traditional optimization methods. As such, the design of efficient algorithms—potentially leveraging advanced mathematical decomposition methods or learning-based optimization—represents a promising research direction for realizing practical PA-assisted AirComp systems.

\textit{Enabling real-time model training using dynamic data}:
Real-time model training in edge AI systems is significantly challenged by the spatial and temporal variability of data generation. 
For instance, in intelligent transportation systems, the data volume generated during daytime—particularly during rush hours—far exceeds that produced at night. Similarly, data generated in densely populated urban centers is much higher than that in suburban areas. Such spatiotemporal variability creates ``data hotspots," where the communication burden for data collection becomes excessively high, leading to network congestion and degraded system performance.

PAs offer new opportunities to alleviate these challenges by adaptively creating LoS links through flexible deployment along dielectric waveguides. By dynamically offloading traffic from congested communication links, PAs can balance the collection of highly dynamic data across time and space. This adaptivity makes PAs particularly well-suited for addressing data tele-traffic hotspot scenarios, thereby supporting more efficient and reliable real-time training in edge AI systems.

\textit{Supporting mobility-aware edge intelligence with stable connectivity}:
In addition to dynamic data patterns, device mobility poses further challenges for real-time edge learning \cite{Mobility_FL}. Mobile devices are subject to rapidly time-varying channel conditions and may even lose connectivity with the server due to physical blockages. Furthermore, frequent handovers between mobile devices and different servers are inevitable in practice. For example, a mobile client may download the latest global model from a specific server, only to move out of its coverage and into another server’s coverage during the local training phase. Such mobility-induced issues can disrupt model updates, increase synchronization delays, and hinder training convergence. Unfortunately, most existing treatises on edge learning largely overlook these challenges, thereby limiting their applicability in dynamic environments.

PAs provide a promising means of mitigating mobility-related issues. By adaptively establishing LoS links, PAs can maintain reliable connectivity for mobile devices despite channel variations, blockages, or server handovers. This capability helps ensure stable model updates even in highly dynamic mobility scenarios, thereby facilitating mobility-aware and real-time edge learning. Designing effective PA activation and activation strategies that adapt to mobility patterns is thus a key research direction to fully exploit this potential.


\vspace{-1em}
\section{Challenges and Future Directions}
\subsection{Interplay Between LLM and PAs for Intelligent NG}
Large language models (LLMs) have recently shown remarkable capabilities in reasoning, contextual understanding, and language generation. Their integration with PAs opens new opportunities to couple semantic intelligence with reconfigurable hardware.
LLMs can interpret heterogeneous inputs—such as network telemetry, user intents, and sensor data—and translate high-level objectives into precise PA control commands. In turn, PAs dynamically adjust pinch activation and waveguide modes to maintain robust LoS links and adapt to rapid-varying traffic or interference.
Conversely, the reconfigurability of PAs supports LLM-driven edge AI by offering low-latency, high-capacity links wherever intensive learning or inference occurs. This mutual synergy bridges intelligent control and adaptive connectivity, enabling scalable and resilient edge AI.
Future work may explore joint design of PA deployment and LLM-based control frameworks to realize distributed intelligence in NG networks.

\vspace{-1em}
\subsection{AI-controlled PAs for Semantic Communications}
Semantic communication (SemCom) shifts transmission from raw bits to compressed semantic information \cite{SemComQin21}. Deep neural encoders and decoders extract and reconstruct semantic features, while AI-controlled PAs can further enhance performance by improving channel conditions that affect semantic similarity. AI models can act as intelligent controllers to optimize PA activation for improved semantic accuracy. Meanwhile, semantic encoders and decoders must adapt to the dynamic propagation environment induced by PA reconfiguration. This coupling motivates joint training of AI controllers and semantic transceivers to co-optimize PA activation and semantic encoding. A further challenge arises from the non-smooth relationship between PA activation position and semantic metrics. Consequently, a promising research direction is the joint optimization of semantic encoders/decoders and PA-activation under a single AI-driven framework for robust, and efficient, context-aware semantic communication.
\vspace{-1em}
\subsection{AI-controlled PAs for ISAC}
Integrated sensing and communication (ISAC) is a cornerstone NG technology that enables simultaneous high-rate transmission and high-accuracy sensing \cite{liu2022integrated}. PAs are particularly suitable for ISAC, as they can improve channel conditions to achieve higher sensing resolution and larger coverage. However, jointly optimizing PA activation for both communication and sensing is challenging due to coupled and conflicting objectives. AI-controlled PAs introduce adaptive intelligence, where deep neural networks dynamically adjust activation positions to balance throughput and sensing accuracy. Future research should pursue unified AI-driven frameworks that jointly design PA control, waveform configuration, and resource allocation to fully exploit AI-controlled PAs for robust and efficient ISAC in NG systems.
\vspace{-1em}
\section{Conclusions} \label{section conclusions}
This article presented the `win-win' collaboration between AI and PAs, which form a mutually beneficial foundation for NG networks. From the AI side, advanced learning techniques, ranging from deep reinforcement learning and GNNs to emerging LLMs, can provide the data-driven intelligence and prompt decision-making required for the high-dimensional, real-time optimization of PA-activation and control. 
From the communication side, PAs fundamentally enhance AI-driven applications by establishing low-cost LoS communication links on demand, which directly alleviate bottlenecks in edge AI networks. The proposed PA-enabled edge AI can guarantee the reliability and latency of federated learning, over-the-air computation, and real-time model training under dynamic traffic or user mobility. 
The interplay between AI and PAs opens several interesting research directions: integrating semantic communication and PA-control under unified AI frameworks, leveraging LLMs for context-aware PA optimization, and co-designing physical layer reconfiguration with high-level learning objectives. These directions will accelerate the realization of self-optimizing, context-aware NG networks, where adaptive PAs and intelligent algorithms jointly deliver the ubiquitous connectivity and distributed intelligence envisioned for future wireless systems.
\vspace{-0.3cm}
\bibliographystyle{IEEEtran}
\bibliography{IEEEfull}
\end{document}